\documentstyle[preprint,aps]{revtex}
\begin{document}
\preprint{nucl-th/9407002}
\draft
\newcommand{\be }{\begin{equation}}
\newcommand{\bea }{\begin{eqnarray}}
\newcommand{\bh }{\begin{displaymath}}
\newcommand{\en }{\end{equation}}
\newcommand{\ena }{\end{eqnarray}}
\newcommand{\eh }{\end{displaymath}}
\title{Friedel Oscillations in Relativistic Nuclear Matter}
\author{Joaquin Diaz Alonso}
\address{D.A.R.C., Observatoire de Paris-Meudon, 92190 Meudon, France}
\author{E. Gallego and A. P\'erez}
\address{Departamento de F\'{\i}sica Te\'orica, Universidad de Valencia
46100 Burjassot (Valencia), Spain}

\maketitle
\begin{abstract}

We calculate the low-momentum N-N effective potential obtained in the OBE
approximation, inside a nuclear plasma at finite temperature, as described by
the relativistic $ \sigma $-$ \omega $ model. We analyze the screening effects
on the attractive part of the potential in the intermediate range as density or
temperature increase. In the long range the potential shows Friedel-like
oscillations instead of the usual exponential damping. These oscillations arise
from the sharp edge of the Fermi surface and should be encountered in any
realistic model of nuclear matter.

\end{abstract}
\vskip 0.5cm

In the usual approaches to the study of nuclear matter near saturation density,
the phenomenological potentials are basic ingredients for the description of
the nuclear interaction in a Schr\"odinger-dynamical framework \cite{FW71}.
Potentials obtained from the exchange of single pseudoscalar, scalar,
pseudovector and vector mesons with different theoretical approaches
\cite{PARIS} lead to satisfactory quantitative predictions of the observed
properties of two-nucleon systems. A simplified description of the main
features of the N-N interaction can be obtained from the exchange of scalar
and vector mesons only \cite{SW86}.

For higher densities (owing to the Pauli principle) relativistic effects become
essential, not only in the description of the interaction itself, but also for
the analysis of the particle dynamics. In this case, the Lagrangian approach
becomes the natural framework for the study of the nuclear plasma \cite{SE93}.
The solution of appropriate Lagrangian models in the relativistic Hartree
approximation (RHA) provides a satisfactory picture of the thermodynamical
behaviour of relativistic nuclear matter at finite temperature \cite{D85}.
Moreover, the relativistic meson propagators in vacuum obtained from this
approach allow for the calculation of the N-N interaction potentials
\cite{MA89}.

Inside nuclear matter, the polarization effects introduce important changes in
the form of the relativistic meson propagators. Consequently, the N-N
interaction potentials inside the plasma are also strongly modified by the
screening. Their behaviour as a function of the thermodynamical state provides
a very explanatory visualization of these effects and suggest the existence of
new collective phenomena.

In this letter we report some results of a study of the N-N interaction
potentials obtained in the one-boson exchange model (OBE) inside symmetric
nuclear matter at finite temperature, and analyze some phenomena related to the
screening.

The Lagrangian model used for this calculation describes the nuclear
interaction in terms of scalar-$ \sigma $ and vector-$ \omega $ mesons
exchanges \cite{SW86}. Although such a simplified model is not able to account
for the whole richness of actual nuclear matter, it gives (when solved in RHA
\cite{CH77}) an acceptable description for its thermodynamical behaviour.
Moreover, as mentioned above, the involved mesons reproduce the main
qualitative features of the nuclear interaction in vacuum. In fact the
fictitious $ \sigma $-meson, which provides the attractive part of the
potential at intermediate range, is introduced as a simple parameterization of
the correlated 2 $\pi$-exchange contribution to the N-N interaction. It has not
been proven that such a parameterization is also possible inside the nuclear
medium. Nevertheless, the behaviour of the screened potential with the
thermodynamical state found here is qualitatively similar, in the intermediate
range, to the one found in a 2 $\pi$-exchange calculation \cite{DU93}.
Moreover, as we shall see, the medium effects on the long-range behaviour of
this potential are mainly dominated by the singularities of the matter
polarization contributions, and are rather independent of the details of the
basic interaction. They should be present in more realistic analysis which take
into account all the relevant meson exchanges.

When solving the model in the RHA, the values of the constants in the
Lagrangian are fixed as follows: The meson and fermion masses are fixed to
their "physical" values $ \mu_{\sigma} \ = \ 550 $ MeV, $ \mu_{\omega} \ = \
783 $ MeV and $ m = 939 $ MeV. For the coupling constants, we choose the values
which lead to a satisfactory fit of the saturation properties in the RHA: $
g_{\sigma}^{2} \ = \ 183.3(\mu_{\sigma}/m)^{2} $ and $g_{\omega}^{2} \ = \
114.7(\mu_{\omega}/m)^{2} $. From these values, saturation is attained at a
Fermi momentum of the nucleon $ P_{f0} = 1.42 fm^{-1} $, with a binding energy
$ E_{b} = -15.46 MeV$.

In going beyond RHA, the analysis of the small perturbations of the fermion
distribution and meson fields around the Hartree equilibrium gives the
expressions for the meson propagation equations inside the plasma, which can
can be written in a matrix form as $ D(k).\Sigma_1 (k) \ = \ 0 $, where $
\Sigma_1 $ is the column matrix of the components of the scalar and vector
perturbing fields, and $D(k)$ is a 5$\times$5-matrix containing the scalar
($\Pi_{\sigma} $), mixing ($\Pi_{\sigma \omega}^ \mu $) and vector ($
\Pi_{\omega} ^{\mu \nu} $) polarization tensors, which are functions of the
thermodynamical state and include the renormalized vacuum contributions
\cite{DP91}. At T=0 they reduce to the one-loop meson polarizations
\cite{MS82}. The propagator matrix for the mixed scalar-vector field inside
matter is given by  \cite{DP91,GDP93} $ G(k) \ =\ -\ D(k)^{-1} $. At
zero density and temperature, the meson fields decouple from each other, and
the components of
this matrix reduce to the one-loop scalar and vector propagators in vacuum.

The first step in the calculation of the screened two-nucleon potential is the
derivation of the relativistic one-boson ($ \sigma \ + \ \omega $) exchange
amplitude diagrams from the propagator. Now, the difference with the
calculation in vacuum \cite{MA89} is that the meson propagators are dressed by
the medium, and mixing between both meson fields appears\cite{DP91}. Therefore,
the usual Feynman rules are slightly modified in the present case: We must
introduce a $\,i\,\Gamma_a\,$ factor at each vertex ($ \Gamma_a\,$ is the
coupling matrix defined as $\,\Gamma_{\mu}=\gamma_{\mu}\,g_{\omega}\,$ for a =
$\mu\,$ from 0 to 3, and $\,\Gamma_4=\,g_{\sigma} \ $), and a dressed boson
propagator matrix $\,i\,G(k)\ $ for each internal boson line. Moreover, in the
present calculation we are interested in the structure of the propagator matrix
on the $ k^0 \ = \ 0 $ axis only, where there are two poles associated to the
"tachyonic" branches, coming from the vacuum polarization terms \cite{DP91}.
Such poles are spurious because they arise at large values of q, where the
point-particle approach fails, and the nucleon structure should be taken into
account. This is done through the introduction of phenomenological monopolar
form factors:
\be
   f_{a} (k)\ =\ (\Lambda_{a}^{2} \ -\
   \mu_{a}^{2})/(\Lambda_{a}^{2} \ -\ k^{2})
   \label{eq:(7)}
\en
($a\ =\ \sigma,\omega$);

at each vertex of the loop and boson-exchange diagrams. This amounts to
multiplying  each squared coupling constant by the
corresponding form factor. With this prescription the spurious "tachyonic"
branches (and the associated poles in the propagator matrix) disappear
\cite{DP91}.

Under these conditions, the relativistic OBE amplitude takes the form

\bea
A \ =\ &
\left\{ \left[ \ \chi^{\dagger}
_{1'} \ \overline{u}(\vec{p'}_1,s'_1)\,\right] \ (i \Gamma)_m \
\left[ \
u(\vec{p}_1,s_1)\ \chi_{1} \ \right] \, \right\} \nonumber \\
 & \cdot (iG(k))^{mn}\
\left\{ \left[ \ \chi^{\dagger}
_{2'} \ \overline{u}(\vec{p'}_2,s'_2)\,\right] \ (i \Gamma)_n \
\left[ \
u(\vec{p}_2,s_2)\ \chi_{2} \ \right] \, \right\}
\label{eq:(8)}
\ena

where $ p_1 $ and $ p_2 $ are the four-momenta for the incoming quasi-nucleon
states and $\,p'_1\,$ and $\,p'_2\,$ correspond to the outgoing quasi-nucleon
states, whereas $\,k=p_1-p'_1=p'_2-p_2\,$ is the transferred four-momentum. The
interacting quasi-nucleons have effective mass M (as given by the RHA
approximation) and spins  $\vec{S}_1 = \frac{1}{2}\vec{\sigma}_1\,$ and
$\vec{S}_2 = \frac{1}{2}\vec{\sigma}_2\,$. The indices m and n run from $\,0\,$
to $\,4\,$. Finally $u(\vec{p},s)$ and $\,\chi \ $ are the Dirac cuadri-spinor
and the isospin wave function, respectively.

The potential (OBEP) is obtained by eliminating the wave functions of the
initial and final states in the amplitude Eq. (\ref{eq:(8)}) taken in the
center-of-mass system. From this expression we obtain the non-relativistic
potential by performing an expansion in the nucleon momenta and keeping only
the second-order terms. We have corrected the OBEP according to the
Blankenblecker-Sugar prescription \cite{ER74}, which includes the requirement
of "minimal relativity" \cite{MA89} as defined by $ V(q',q)\ =\
(M/E_{q'})^{1/2}\,V_{OBEP}(q',q)\,(M/E_q)^{1/2} $ where $q$ and $q'$ are the
CMS initial and final momenta of the nucleons, respectively, $E_q\ =\ \sqrt{M^2
+ q^2}$, and $E_{q'}\ =\ \sqrt{M^2 + {q'}^2}$. Moreover, the above prescription
implies taking the static ($k^0 = 0 $) limit. After Fourier transformation one
obtains the potential in coordinate space:

\be
V(\vec{r}) \ =\ V_c(r) \,-\,\frac{1}{2} ( \nabla^{2} V_2(\vec{r})+
V_2(\vec{r}) \nabla^{2}) \,+\,V_{LS}(r) \, \vec{L} \cdot \vec{S}
\,+\,V_{SS}(r) \, \vec{\sigma}_1  \cdot \vec{\sigma}_2  \,+\,
V_T(r) \,S_{12}
\label{eq:(10)}
\en

where $\,\vec{L}=\vec{r} \wedge \vec{p}\,$ is the orbital kinetic momentum,
$\,\vec{S}= (1/2)(\vec{\sigma}_1\,+\,\vec{\sigma}_2)$ is the total spin
operator and $\ \,S_{12}= \frac{3}{r^{2}}  (\vec{\sigma}_1 \cdot \vec{x})
\,(\vec{\sigma}_2 \cdot \vec{x}) \,-\,(\,  \vec{\sigma}_1 \cdot \vec{\sigma}_2
\,)$ is the tensor operator.

The second term in Eq.(\ref{eq:(10)}) is a non-local component which gives a
small contribution to the potential. We shall omit here the study of this
component. The other terms (central, spin-spin, spin-orbit and tensor
components) are now functions of the interparticle distance, as well as of the
thermodynamical state of the plasma (density and temperature). We must
emphasize that the non-relativistic limit concerns the dynamics of the two
interacting nucleons and the neglecting of retardation effects. However, the
description of the thermodynamical state of the plasma in the RHA remains fully
relativistic. Also, no low-q approximation has been done for the meson
propagators.

In the analysis of the potential, the values of the coupling constants and
cut-off parameters in the form factors have been fixed in order to fit the
deuteron and low-energy phenomenology data \cite{MA89}:

\be
g_{\sigma}^2=8.7171 \ \ \ \ \  \ g_{\omega}^2=25 \ \ \ \ \ \Lambda_
{\sigma}=2.0
\,GeV \ \ \ \ \ \ \Lambda_{\omega}=1.4 \, GeV
\label{const}
\en

However, in calculating the underlying thermodynamical state, the coupling
constants have been fixed to the above mentioned values which fit saturation in
RHA.

Figure (1) is a picture of the central component of the potential at $ T = 0 $,
at saturation density (solid line) and 2.4 times these density (dashed-dotted
line), for symmetric nuclear matter ($P_f$ is the Fermi momentum of the
nucleons). The same component, calculated at zero
density, with ($V_{pol}$) and without ($V_{vac}$) vacuum polarization
contributions has also been plotted. The very short-range region ($ r < 0.6 fm
$) must be discarded because the potential behaviour is dominated there by
large-q values, and the non-relativistic (low-q) limit breaks down
\cite{GDP93}. Beyond this distance, a hard-core and a potential well appear
both in vacuum and at finite density. At zero density, the vacuum polarization
effects enhance the slope of the hard-core and increase the depth of the
potential well by an important amount. The slope of the repulsive-vector Yukawa
component is strongly raised by the vacuum polarization, whereas the
attractive-scalar component remains nearly unaffected by these effects. This
explains the observed behaviour in vacuum. The matter polarization effects
increase the range of the vector component, and therefore the depth of the well
is reduced as density grows. It disappears shortly above the saturation
density \cite{GDP93}. Similar results are obtained if the
attractive $\sigma$ contribution to the N-N potential is replaced by the 2
$\pi$-exchange \cite{DU93}.

At larger distances, the screening effects of the medium introduce important
qualitative new features. Whereas in vacuum the potential shows an exponential
damping with the distance, an oscillatory behaviour appears at finite density,
whose amplitude is damped as an integer power of the distance. These are
Friedel oscillations, similar to those encountered in many low-temperature
Fermi systems \cite{FR52}. Mathematically, these oscillations arise from the
fact that the matter polarization contributions to the screened meson
propagators show singularities in their derivatives at $ q = 2P_{f} $ (Kohn
singularities \cite{KO59}). After Fourier transformation, such singularities
introduce oscillations in the r-space potential. More detailed analytical
calculations \cite{DPS89},\cite{DGPS93} show that, in this region, the
potential can be decomposed into $\sigma$ and $\omega$ Yukawa-like components
(which dominate at shorter distances), and this Friedel-like component, which
determines the large-r features and is long-ranged and oscillatory. From a
physical point of view, the Kohn singularity is associated to the sharp
character of the Fermi surface at $ T = 0 $ and is rather independent on the
details of the interaction. Consequently, the long-range oscillations of the
screened interparticle potential must be an universal feature of interacting
degenerate Fermi systems. Indeed, this has been found for degenerate
non-relativistic \cite{REN68} and relativistic \cite{SI85} electromagnetic
plasmas, for a QCD plasma \cite{KA88}, and for the screened one-pion exchange
potential in a relativistic nuclear plasma \cite{DPS89}.

In the calculation of Ref. \cite{DU93}, where the 2 $\pi$-exchange is
explicitly considered, no Friedel oscillations have been found. In fact, this
paper was concerned with the intermediate range of the screened N-N potential,
and the complete effects of the matter polarizations were not included in the
meson propagators. Such effects go beyond the effective meson mass variation
with density, considered by the authors. Therefore, it is not surprising that
they do not obtain oscillations in the long range.

As a test of the small sensitivity of the Friedel behaviour to the details of
the interaction, we have plotted in Fig.(1) the central component of the
potential at saturation density and $ T = 0 $ (dotted line), now using for the
potential the values of the coupling constants which fit saturation in the RHA,
which are very different from the ones used in the previous calculation for the
same thermodynamical state. We observe important differences in the
intermediate region in both cases but, in the long range, the oscillations of
the potential are only slightly affected by the changes in the intensity of the
couplings.

In Fig. (2) the spin-spin component of the potential is plotted at $ T = 0 $,
in vacuum ($ P_f = 0 $), at saturation density ($ P_f/m = 0.3 $), and 2.4 times
saturation density ($ P_f/m = 0.4 $). A glance to this drawing shows also a
Friedel-like  oscillatory behaviour whose amplitude increases with density.
(The spin-orbit and tensor components of the potential show also a similar
behaviour: in all cases the amplitude of the oscillations increase as density
grows \cite{GDP93}). At saturation density, the most important effect in the
long-range appears on the central potential component, where the amplitude of
the first oscillation reaches $2 MeV$ for a distance of around $2fm$.
Nevertheless, for higher densities the oscillations of the spin-spin component
reach comparable maxima. Indeed, for $ P_f/m = 0.4 $ the amplitude of the first
oscillation in the central component is $\sim 8 MeV$ at $ r_{max}\ \approx
1.5fm $ (see Fig.(1)), whereas the spin-spin component reaches $10 MeV$ at the
same distance (see Fig.(2)). The position of the first maximum of the
oscillations, $\,r_{max}\,$, is related to the Fermi momentum through
\cite{DPS89} $ Pp_f . r_{max} \approx \pi $. The ratio between the position of
the first maximum of the oscillation and the mean interparticle distance
$d=(3\pi^{2} / 2 P_f^{\,3})^{1/3}\,$ is nearly constant with density ($
r_{max}/d \approx 1.28 $).

At finite temperature, the slope of the hard core is slightly reduced, as
showed in Fig.(3), where the central component has been plotted at saturation
density and various temperatures. In the intermediate range, the effects of
temperature on the central potential reduce the depth of the well. The well
disappears for temperatures beyond 40 MeV. This can be interpreted in terms of
the modifications introduced on the Yukawa-like components by the temperature
in this region. As temperature increases, the ranges and intensities of these
Yukawa components are modified in such a way that in the balance, the
vector-repulsive part becomes dominant and the well disappears. (see Refs.
\cite{DPS89} and \cite{DGPS93} for a more detailed analytical study).

Concerning the long-range behaviour of the potential, as temperature increases
the Fermi distribution function becomes smooth and the Kohn singularity (and
the associated oscillations) disappears in all the components of the potential.
This is indeed observed in Fig. (3). The amplitudes of the oscillations
decrease with temperature. As can be checked analytically \cite{DPS89,DGPS93}
this amplitude at a fixed distance and constant density is exponentially damped
with temperature. The oscillations in all the components disappear for
temperatures between $40-80 MeV$. Beyond this, the potential becomes
exponentially damped with the distance.

\acknowledgments

This work has been supported in part by the spanish CICYT, Grant AEN 93-0234
and DGICYT Grant PB91-0648.

\begin{figure}
\caption{Central component of the potential at $ T = 0 $ at zero density with
($ V_{pol} $) and without ($ V_{vac} $) vacuum polarization contributions at
saturation density ($ P_f /m\ =\ 0.3 $) and 2.4 times saturation density
($ P_f /m\ =\ 0.4 $). The dotted line (with $ P_f /m\ =\ 0.3 $) corresponds to
the central component of the potential at saturation density obtained from
different values of the model parameters.}
\end{figure}

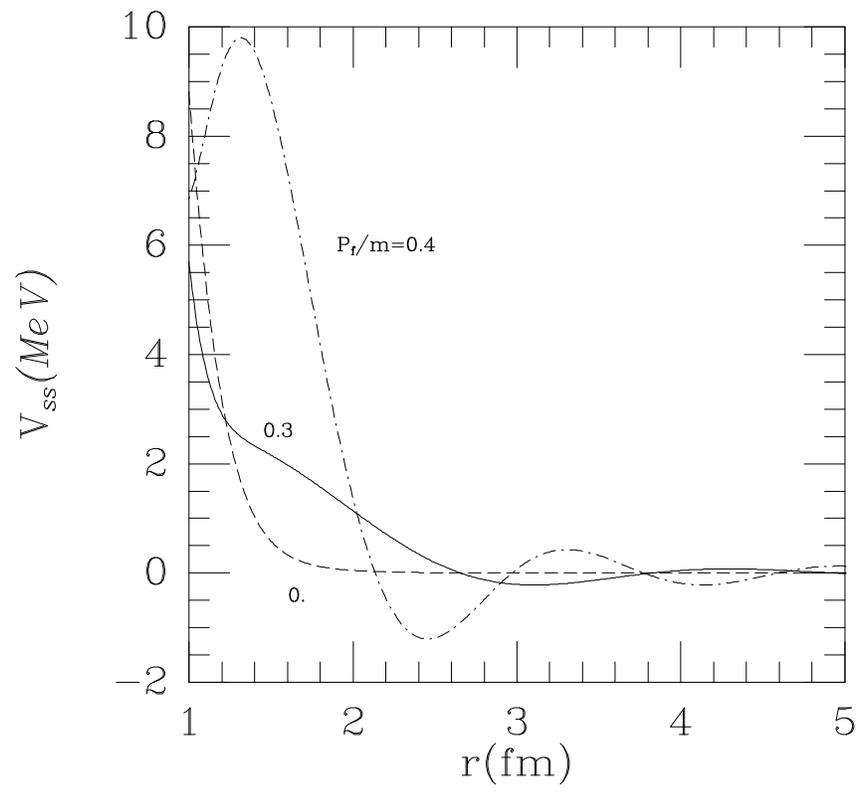
\begin{figure}
\caption{Spin-spin component of the potential at $ T = 0 $
for the same values of the Fermi momentum as in Fig.1.}
\end{figure}

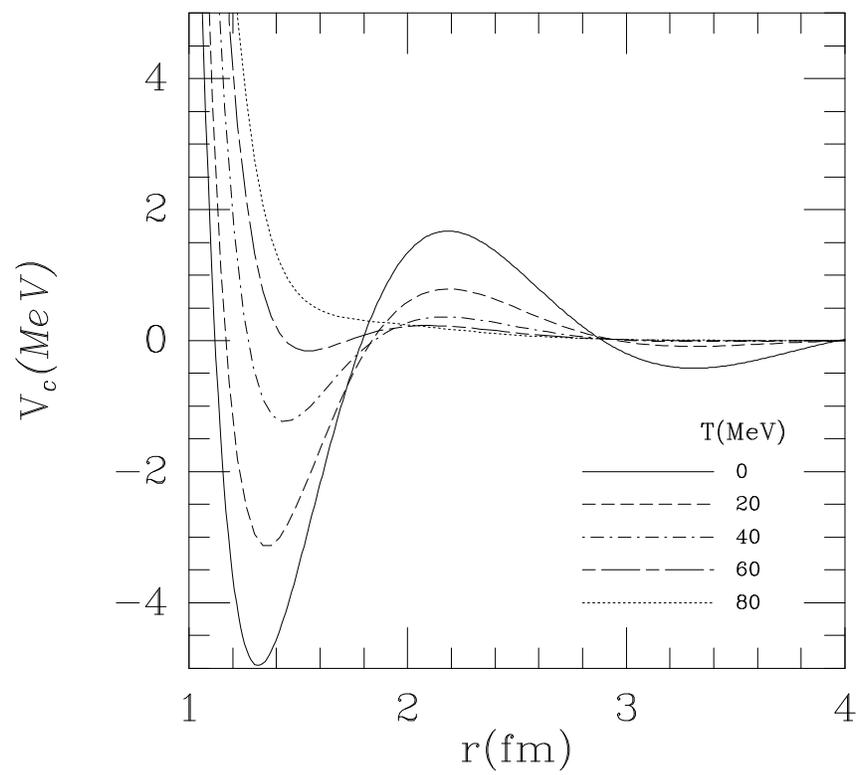
\begin{figure}
\caption{Central component of the potential at saturation density for
different temperatures.}
\end{figure}


\begin{references}

\bibitem{FW71} {A.L. Fetter, J.D. Walecka : "Quantum Theory of Many-Particle
Systems". Mc-Graw-Hill (1971).}

\bibitem{PARIS} {M. Lacombe, B. Loiseau, J.M. Richard, R. Vinh Mau, J. Cote,
P. Pires, R. de Tourreil : Phys. Rev. {\bf C21}, (1980), 861.
\newline R. Machleidt, K. Holinde, C. Elster : Phys. Rep. {\bf 149}, (1987),
1.}

\bibitem{SW86}{B.D. Serot, J.D. Walecka: "The Relativistic Nuclear Many-Body
Problem." Adv. Nucl. Phys. {\bf Vol.16}, J.W. Negele and E. Vogt Ed.
Plenum Press, New York (1986).
\newline J.D. Walecka : Ann. Phys. {\bf 83}, (1974), 491.}

\bibitem{SE93} {B.D. Serot : Rep. Prog. Phys. {\bf 55}, (1992), 1855.}

\bibitem{D85}{J. Diaz Alonso: Phys. Rev. {\bf D31}, (1985), 1315}

\bibitem{MA89}{R. Machleidt: "The Meson Theory of Nuclear Forces and
Nuclear Structure." Adv. in Nucl. Phys. {\bf Vol.19}, J.W. Negele and E. Vogt
Ed. Plenum Press, New York (1989)}

\bibitem{CH77} {S.A. Chin : Ann. of Phys. {\bf 108}, (1977), 301.}

\bibitem{DU93} {J.W. Durso, H-C. Kim, J. Wambach: Phys. Lett. {\bf B 298},
(1993), 267.}

\bibitem{DP91}{J. Diaz Alonso, A. Perez Canyellas: Nucl. Phys. {\bf A 526},
        (1991), 623}

\bibitem{MS82}{T. Matsui, B.D. Serot: Ann. of Phys. {\bf 144}, N1, (1982), 107}

\bibitem{GDP93}{E. Gallego, J. Diaz Alonso, A. Perez Canyellas:
     "Screening Effects on the N-N Potential in Relativistic Nuclear
      Matter." In preparation.}

\bibitem{ER74}{K. Erkelenz: Phys. Rep. {\bf 13 C}, (1974), 191.
\newline Ch. W. Wong: "Topics in Nuclear Physics I", Lecture Notes
 in Physics, 144, Springer-Verlag (1981)}

\bibitem{FR52}{J. Friedel: Phyl. Mag. {\bf 43}, (1952), 153.;
\newline Nuovo Cimento, {\bf 7}, (1958), Suppl.2, 287.}

\bibitem{KO59}{W. Kohn: Phys. Rev. Lett. {\bf 2}, (1959), 393.
\newline J.S. Langer, S.H. Vosko: J. Phys. Chem. Solids, {\bf 12},
(1960), 196.}

\bibitem{DPS89}{J. Diaz Alonso, A. Perez Canyellas, H. Sivak: Nucl. Phys.
    {\bf A205}, (1989), 695}

\bibitem{REN68}{M.E. Rensik: Phys. Rev. {\bf 174}, (1968), 744.
\newline N.J. Horing: Phys. Rev. {\bf 186}, (1969), 434}

\bibitem{SI85} {H. Sivak : Physica, {\bf A129}, (1985), 408.}

\bibitem{KA88} {J. Kapusta, T. Toimela : Phys. Rev. {\bf D37}, (1988), 3731.}

\bibitem{DGPS93}{J. Diaz Alonso, E. Gallego, A. Perez Canyellas, H.Sivak:
     "Linear Response and Friedel Oscillations of Meson Fields in
      Relativistic Nuclear Matter." To be published.}

\end{references}
\end{document}